\documentclass[aps,prd,email,twocolumn,showpacs,showkeys,preprintnumbers,amsmath,amssymb,nofootinbib]{revtex4}

\usepackage{amsmath,latexsym}
\def\[{\left\lbrack}
\def\]{\right\rbrack}

\def\({\left(}
\def\){\right)}
\newcommand{\be}{\begin{equation}}
\newcommand{\ee}{\end{equation}}
\newcommand{\ea}{\end{eqnarray}}
\newcommand{\ba}{\begin{eqnarray}}

\def\ni{\noindent}
\def\no{\nonumber \\}

\begin{document}

\title{Faddeev-Jackiw analysis for the charged compressible fluid in a higher-derivative electromagnetic field background}

\author{Albert C. R. Mendes}\email{albert@fisica.ufjf.br}
\affiliation{Departamento de F\'{i}sica, Universidade Federal de Juiz de Fora, 36036-330, Juiz de Fora - MG, Brazil}
\author{Everton M. C. Abreu}\email{evertonabreu@ufrrj.br}
\affiliation{Grupo de F\' isica Te\'orica e Matem\'atica F\' isica, Departamento de F\'{i}sica, Universidade Federal Rural do Rio de Janeiro, 23890-971, Serop\'edica - RJ, Brazil}
\affiliation{Departamento de F\'{i}sica, Universidade Federal de Juiz de Fora, 36036-330, Juiz de Fora - MG, Brazil}
\author{Jorge Ananias Neto}\email{jorge@fisica.ufjf.br}
\author{Flavio I. Takakura}\email{flavio@fisica.ufjf.br}
\affiliation{Departamento de F\'{i}sica, Universidade Federal de Juiz de Fora, 36036-330, Juiz de Fora - MG, Brazil}


\date{\today}



\begin{abstract}
\ni In the present paper we will discuss the Faddeev-Jackiw symplectic approach in the analysis of a charged compressible fluid immersed in a higher-derivative electromagnetic field theory.
We have obtained the full set of constraints directly from the zero-mode eigenvectors.
Besides, we have computed the Dirac brackets for the dynamic variables of the compressible fluid. Finally, as a result of the coupling between the charged compressible fluid and the electromagnetic field we have calculated two Dirac brackets between the fluid and electromagnetic fields, which are both zero when there is no coupling between them.
\end{abstract}

\maketitle

\section{Introduction}

The search for a connection between fluid dynamics and electromagnetism is an old concept and it has played a crucial role in the development of the Maxwell equations \cite{Whittaker,Everitt}. Thomson applied analogous formulations connecting electrostatics, heat transfer and elasticity of solids, that later lead Maxwell to formulate his theory of electricity and magnetism \cite{Whittaker}. This analogy was first applied to the set of Maxwell equations concerning fluid dynamics in the early 1962 \cite{Logan} only to the case of the one-dimensional Rayleigh problem.   Recently, the generalization of the Maxwell set of fluid equations was introduced in terms of an incompressible flow, particularly with interest in turbulent flow \cite{Marmanis}. Even more recently, the fluid Maxwell equations were generalized to the compressible flow case \cite{Kambe}. Besides, other generalizations of fluid dynamics have been constructed proposing noncommutative, non-Abelian and supersymmetric formulations, to mention a few \cite{jackiw}.

In \cite{albert1}, some of us have introduced a Lagrangian description for the compressible fluid together with the scenario where a charged fluid  was immersed in an electromagnetic field.   The interaction between them from the Lagrangian density was discussed.  This analogy has been explored in the literature with applications in  quark-gluons plasma (QGP) \cite{jackiw, heinz, choquet, hk, patten, bi} which is a dense liquid that flows with very little viscosity almost being an ideal  fluid.

Having said that, we can consider this work as part of a sequence of other ones from these authors upon the analysis of this mentioned analogy between the structure of the fluid dynamics and electrodynamics \cite{albert0,albert1,albert2}.  The purpose of the present paper is to analyze the Lagrangian density which describes the charged compressible fluid immersed in an electromagnetic field, obtained in  \cite{albert1}, from the point of view of the Faddeev-Jackiw method \cite{FJ} applied to this model.

The Faddeev-Jackiw (FJ)  \cite{FJ} method is a symplectic description of constrained quantization, where the degrees of freedom are identified by means of the so-called symplectic variables. The essential point  of the FJ method is to make the system into a first order Lagrangian with some auxiliary fields, but the method does not depend on how the auxiliary fields are introduced to make the first order Lagrangian. It was applied recently in non-Abelian theories \cite{bp}.

The work is organized in such a way that in section 2 we have reviewed briefly  the FJ method. In section 3 we have analyzed the theory via the Faddeev-Jackiw method and finally in the last section we present the conclusions.

\section{Faddeev-Jackiw formalism }


We will begin with a first-order time derivative Lagrangian, which arises from a standard second-order one with auxiliary fields. 
The first step is to construct the symplectic Lagrangian 
\begin{equation}
\mathcal{L}=a_{i}\left( \xi \right) \dot{\xi}^{i}-\mathcal{V}\left( \xi
\right) ,  \label{eq 1.0}
\end{equation}%

\ni where $a_{i}$ are the arbitrary one-form components and $i=1,...,N$. 
Since the first-order system is constructed through a closed two-form, if it is non-degenerated, it defines a symplectic framework on the phase space, which is
described by the coordinates $\xi _{i}$. 
Besides, if this two-form is singular, with constant rank, it is defined as a pre-symplectic two-form.
Hence, considering the components, the symplectic form can be defined by
\begin{equation}
f_{ij}=\frac{\partial }{\partial \xi ^{i}}a_{j}\left( \xi \right) -\frac{%
\partial }{\partial \xi ^{j}}a_{i}\left( \xi \right) \,\,,  \label{eq 1.1}
\end{equation}

\ni and the equations of motion are
\be \label{eq 1.2}
f_{ij}\dot{\xi}^{j}=\frac{\partial }{\partial \xi ^{i}}\mathcal{V}\left( \xi
\right) \,\,,  
\ee 

\ni where the two-form $f_{ij}$ can be either singular or nonsingular.  In this last case it has an inverse $f^{ij}$
\begin{equation}
\dot{\xi}^{i}=f^{ij}\frac{\partial }{\partial \xi ^{j}}\mathcal{V}\left( \xi
\right) ,  \label{eq 1.3}
\end{equation}

\ni where we have that $\left\{ \xi ^{i},\xi ^{j}\right\} =f^{ij} $.
To consider a constrained system described by \eqref{eq 1.0}, it means that the symplectic matrix is singular.  And the constraints
of the system have to be determined, of course. Consider that the rank of $%
f_{ij}$ is $2n$.  In this case we have $N-2n=M$ zero-mode vectors $\mathbf{\nu}^{\alpha
} $, $\alpha =1,...,M$. The system is then constrained through $M$ equations with
no time-derivatives. 
We will have constraints that reduce
the degrees of freedom's number. Hence, multiplying \eqref{eq
1.2} by the (left) zero-modes $\mathbf{\nu}^{\alpha }$ of $f_{ij}$ we have the
(symplectic) constraints with the structure of algebraic relations
\begin{equation}
\Omega ^{\alpha }\equiv \mathbf{\nu}_{i}^{\alpha }\frac{\partial }{\partial
\xi ^{i}}\mathcal{V}\left( \xi \right) =0\,\,.  \label{eq 1.5}
\end{equation}%

\ni So, we can construct the first-iterated Lagrangian by including the 
corresponding Lagrange multipliers relative to the obtained constraints
\begin{equation}
\mathcal{L}=a_{i}^{\left( 1\right) }\left( \xi \right) \dot{\xi}^{i}+\Omega
^{\alpha }\lambda _{\alpha }-\mathcal{V}^{\left( 1\right) }\left( \xi
\right) .
\end{equation}

\ni The Lagrange multipliers $\lambda $ can be considered as the
symplectic variables which can increase the symplectic variables set. This
move reduces the number of $\xi $'s.  After that, the procedure can be entirely
repeated until all the constraints can be eliminated and the
completely reduced, unconstrained and canonical system remains. But notice that in the case of gauge theories, we have no new constraint through
the zero-mode.
And the symplectic matrix remains singular. Hence, we can consider
mandatory to introduce gauge condition(s) to highlight the singularity. In this way the
procedure can be finished in terms of the original variables.  And the basic brackets can be determined.


\section{Faddeev-Jackiw analysis for the charged compressible fluid immersed in an electromagnetic field}

The effective Lagrangian density which describes the charged compressible fluid immersed in an electromagnetic field is defined, valid for each species 
$(\epsilon)$, by
\be \label{01}
{\cal L} = -{1\over4}T_{(\epsilon)}^{\mu\nu}T^{(\epsilon)}_{\mu\nu} -{(1+g^2)\over 4} F^{\mu\nu}F_{\mu\nu} -{g\over2}T_{(\epsilon)}^{\mu\nu}F_{\mu\nu}\,\,,
\ee

\ni where $T^{(\epsilon)}_{\mu\nu} =\partial_\mu U^{(\epsilon)}_{\nu} -\partial_\nu U^{(\epsilon)}_{\mu}$ is the strength tensor of the fluid, the four-vector potential $U^{(\epsilon)}_\mu \equiv (U^{\epsilon}_0, \vec{U}^{\epsilon})$ - $U^{\epsilon}_0$ is the energy function and $\vec {U}^{\epsilon}$ is the average velocity field \cite{albert1} - and $F_{\mu\nu} = \partial_\mu A_{\nu} -\partial_\nu A_{\mu}$ is the strength tensor for the electromagnetic field.  The spacetime metric elements are $\eta_{\mu\nu}=(-+++)$. The coupling constant is $g=e_{\epsilon}/m_{\epsilon}$, where $e_{\epsilon}$ is the charge and $m_{\epsilon}$ is the mass of the charge. Note that, when $g=0$ we have  two uncoupled theories.
The Euler-Lagrange equations of motion are
\be\label{02.1}
(1+g^2)\partial_\mu F^{\mu\nu} + g\partial_\mu T_{(\epsilon)}^{\mu\nu} =0,
\ee
and it is easy to see that (\ref{01}) is invariant under the gauge transformations,
$A_\mu \rightarrow A_\mu +\partial_\mu \Lambda$,
for the electromagnetic fields, and
$U^{(\epsilon)}_\mu \rightarrow U^{(\epsilon)}_\mu +\partial_\mu \Lambda$,
for the compressible fluid field. In terms of the potentials, $U^{(\epsilon)}_\alpha$ and $A_\alpha$, the above equation reads
\be\label{02.4}
(1+g^2)\[ {\Box} A_{\mu} - \partial_{\mu} \partial_{\nu} A^{\nu} \] + g\[ {\Box} U^{(\epsilon)}_{\mu} -\partial_{\mu}\partial_{\nu} U_{(\epsilon)}^{\nu} \]=0.
\ee
From now on, for simplicity, we will not use the species index, and much of what follows is true for each species. In our model, the symmetric energy-momentum tensor is given by
\ba\label{02.5}
&&\Theta^{\alpha\beta} \no
&=& (1+g^2)\left[ \eta^{\alpha\mu}F_{\mu\lambda}F^{\lambda\beta} + {1\over 4}\eta^{\alpha\beta}F_{\mu\lambda}F^{\mu\lambda}\right] \no
&+& \left[ \eta^{\alpha\mu}T_{\mu\lambda}T^{\lambda\beta} + {1\over 4}\eta^{\alpha\beta}T_{\mu\lambda}T^{\mu\lambda}\right] \\ 
&+&g\left[ \eta^{\alpha\mu}T_{\mu\gamma}F^{\gamma\beta} + \eta^{\alpha\mu}F_{\mu\gamma}T^{\gamma\beta} +{1\over 2}\eta^{\alpha\beta}T^{\mu\lambda}F^{\mu\lambda}\right] \nonumber
\ea
and it follows directly that
\ba\label{02.6}
&&\Theta^{00} = {1\over 2}\( {\vec l}^{\,\,2} + {\vec\omega}^2 \) + {(1+g^2)\over 2}\( {\vec E}^{\,\,2} - {\vec B}^2 \)  \no
&&+\,g\,{\vec l} \cdot \vec{E} + \,g\,{\vec \omega}\cdot\vec{B}
\ea 

\ni which is the energy of the model, where the first term in (\ref{02.6}) is the energy of the fluid and the second one is the energy of the electromagnetic field. The last two terms are the contributions of the interaction between the two fields.

As we said before, in this paper we want to discuss the Faddeev-Jackiw methodology \cite{FJ} applied in the analysis of a higher-derivative theory which, in this case, have the higher-derivative in the Maxwell sector.  So, rewriting (\ref{01}) in the form
\be\label{02}
{\cal L} = -{1\over4}T^{\mu\nu}T_{\mu\nu} -{(1+g^2)\over 4} F^{\mu\nu}F_{\mu\nu} -g U_{\mu}\partial_{\nu}F_{\mu\nu},
\ee

\ni we can   introduce another set of canonical pair $(\Sigma^{\mu}\equiv \partial_0 A^{\mu}, \phi)$ in order to have a correct extended phase space in order to proceed with the canonical analysis. Therefore, we have that
\ba\label{03}
&&{\cal L} ={1\over 2}(\dot{\vec U} -\nabla U_0 )^2 + {1\over2}(\nabla \times \vec{U})^2 \no
&&+ {1\over2}(1+g^2)(\vec \Sigma -\nabla A_0)^2 +{1\over 2}(1+g^2)(\nabla\times\vec A)^2 \nonumber\\ 
&&-g\vec{U}\cdot(\nabla\Sigma_0 -\dot{\vec\Sigma}) -gU_0 (\nabla.\vec\Sigma -\nabla^2 A_0 ) \no
&&-\,g\vec{U} \cdot (\nabla\times\nabla\times \vec A)\,\,,
\ea

\ni and to write  a first order Lagrangian, we will use an auxiliary field, which is chosen to be the canonical momentum due to an algebraic simplification.  In this case, we have a set of canonical pairs $(U_\mu , p_\mu )$, $(A_\mu , \pi_\mu )$ and $(\Sigma_\mu , \phi_\mu )$ and we have directly that
\ba\label{04}
p_\mu &=& {{\partial {\cal L}}\over{\partial(\partial_0 U_\mu)}}\:\: ,\,\,\,\,
\phi_\mu = {{\partial {\cal L}}\over \partial{(\partial_0 \Sigma_\mu)}}  \,\,,\\
\pi_\mu &=& {{\partial {\cal L}}\over \partial{(\Sigma_\mu)}} -\partial_0 {{\partial {\cal L}}\over \partial{(\partial_0 \Sigma_\mu)}} -2\partial_k {{\partial {\cal L}}\over \partial{(\partial_k \Sigma_\mu)}} \,\,, \nonumber
\ea

\ni which results in the following expressions
\ba\label{05}
p_\mu &=& T_{\mu 0}\quad , \quad 
\phi_\mu = g\eta_{\mu k}U_k \,\,,\\
\pi_\mu&=&(1+g^2)F_{\mu 0} -g\eta_{\mu k}T_{0k} +g\eta_{\mu 0}\partial_k U_k \,\,.\nonumber
\ea

\ni Therefore, making use of the equation of motion for the canonical momenta associated with the fields $U_\mu$, $A_\mu$ and $\Sigma_\mu$, we have
\be\label{06}
{\cal L}^{(0)} = -\vec{p}\cdot\dot{\vec{U}} +\vec{\phi}\cdot\dot{\vec \Sigma} +\pi_\mu \dot{A}^{\mu} - V^{(0)} \,\,,
\ee

\ni where the potential density is
\ba\label{07}
&&V^{(0)}= \pi_{\mu}\Sigma^{\mu} -{1\over2}{\vec p} ^2 -\vec p \cdot\nabla U_0 - {1\over2}(\nabla \times \vec{U})^2 \\
&-& {1\over2}(1+g^2)(\vec \Sigma -\nabla A_0)^2 -{1\over 2}(1+g^2)(\nabla\times\vec A)^2 \nonumber\\ 
&+&\!\!\!\vec{\phi}\cdot \nabla\Sigma_0 +gU_0 (\nabla\cdot\vec\Sigma -\nabla^2 A_0 ) +\vec{\phi} \cdot (\nabla\times\nabla\times \vec A). \nonumber
\ea

The initial set of symplectic variables defining the extended space is given by the set $\xi^{(0)}=(U_k, p_k, U_0; A_k, \pi_k, A_0, \pi_0; \Sigma_k, \pi_k, \Sigma_0 )$, and the corresponding canonical non-zero one-form  is
\be\label{08}
{}^{U}a^{(0)}_k =-p_k ; \, {}^{\Sigma}a^{(0)}_k =\phi_k;\, {}^{A}a^{(0)}_k =-\pi_k;\,{}^{A_0}a^{(0)} =\pi_0. \nonumber
\ee

\ni Using this result in the symplectic two-form matrix $f^{(0)}$ we have that
\be\label{09}
f^{(0)}_{ij}(\vec x, \vec y) =
\begin{pmatrix}
{\bf F}_{ij} & {\bf 0}_{3\times4} & {\bf 0}_{3\times3} \cr {\bf 0}_{4\times3} & {\bf M}_{ij} & {\bf 0}_{4\times 3}\cr {\bf 0}_{3\times3} & {\bf 0}_{3\times4} & {\bf C}_{ij} \cr
\end{pmatrix} \delta(\vec x -\vec y)
\ee
with
\ba\label{10}
&&{\bf F}_{ij}=
\begin{pmatrix}
0 & \delta_{ij} & 0 \cr -\delta_{ij} & 0 & 0 \cr 0 & 0 & 0 \cr
\end{pmatrix} ,
{\bf M}_{ij}=
\begin{pmatrix}
0 & \delta_{ij} & 0 & 0 \cr -\delta_{ij} & 0 & 0 & 0 \cr 0 & 0 & 0 & -1 \cr 0 & 0 & 1 & 0 \cr
\end{pmatrix}, \no
&&{\bf C}_{ij}=
\begin{pmatrix}
0 & -\delta_{ij} & 0 \cr \delta_{ij} & 0 & 0 \cr 0 & 0 & 0 \cr
\end{pmatrix}\,\,,
\ea

\ni where we can note  that the matrix $f^{(0)}_{ij}$ is singular, which  means that there are constraint and it has two zero-mode
${}^{1}\nu_{\gamma} \equiv ({\bf 0},{\bf 0},\nu^{U_0},{\bf 0},{\bf 0},0,0,{\bf 0},{\bf 0},0)$
and
${}^{2}\nu_{\gamma} \equiv ({\bf 0},{\bf 0},{\bf 0},{\bf 0},{\bf 0},0,0,{\bf 0},{\bf 0},\nu^{\Sigma_0})$
where $\nu^{U_0}$ and $\nu^{\Sigma_0}$ are arbitrary function. From this two zero-mode, we have the following constraints
\ba\label{13}
&&{}^{1}\Omega^{(0)}=\int d^3\vec x \,\nu^{U_0}(\vec x){{\delta}\over{\delta U_0(\vec x)}}\int d^3\vec y \,V^{(0)}(\vec y) \\
&=&\!\!\!\!\!\int d^3\vec x\,\nu^{U_0}(\vec x)\[ \nabla \cdot \vec p(\vec x) + g(\nabla \cdot\vec\Sigma(\vec x) -\nabla^2 A_0(\vec x) )\]  \nonumber \\
&=&0
\ea
\ba\label{14}
\mbox{and}&&\!\!\!\!\!\!\!\!\!\!\quad {}^{2}\Omega=\int d^3\vec x \,\nu^{\Sigma_0}(\vec x){{\delta}\over{\delta \Sigma_0(\vec x)}}\int d^3\vec y \,V^{(0)}(\vec y) \nonumber\\
&=&\int d^3\vec x\,\nu^{\Sigma_0}(\vec x)\[ \pi_0(\vec x) -\nabla.\phi(\vec x) \] =0\,\,.
\ea

\ni Since $\nu^{U_0}$ and $\nu^{\Sigma_0}$ are arbitrary functions, we obtain the constraints
\be\label{15}
{}^{1}\Omega= \nabla \cdot \vec p + g(\nabla \cdot\vec\Sigma -\nabla^2 A_0 ) =0
\ee

\ni and
\be\label{16}
{}^{2}\Omega=\pi_0 -\nabla\cdot\phi =0\,\,.
\ee

According to the symplectic algorithm, the constraints (\ref{15}) and (\ref{16}) are introduced in the Lagrangian density by using the Lagrangian multipliers. Thus, the first iterated Lagrangian density is written as  
\be\label{17}
{\cal L}^{(1)} = -\vec{p}\cdot\dot{\vec{U}} +\vec{\phi}\cdot\dot{\vec \Sigma} +\pi_\mu \dot{A}^{\mu} + \dot{\lambda}_1 {}^{1}\Omega + \dot{\lambda}_2 {}^{2}\Omega  - V^{(1)}\,\,,
\ee

\ni where $\lambda_1$ and $\lambda_2$ are the Lagrangian multipliers, and the first iterated symplectic potential density is
\ba\label{18}
&&V^{(1)}= V^{(0)}\Big|_{{}^{1}\Omega=0, {}^{2}\Omega=0} \no
&&= -\,\vec{\pi}\cdot\vec{\Sigma} -{1\over2}{\vec p}^{\,\,\,2}  - {1\over2}(\nabla \times \vec{U})^2 \no
&&- \,{1\over2}(1+g^2)(\vec \Sigma -\nabla A_0)^2 -{1\over 2}(1+g^2)(\nabla\times\vec A)^2 \nonumber\\ 
&&+\,\vec{\phi} \cdot (\nabla\times\nabla\times \vec A)\,\,.
\ea

\ni It should be noted that when the constraints ${}^{1}\Omega$ and ${}^{2}\Omega$ are imposed the dependence in $U_0$ and $\Sigma_0$ disappears, once the terms in $U_0$ and $\Sigma_0$ were incorporated in the term introduced to the Kinetic part, which was done by redefining the Lagrange multipliers.

From the above Lagrangian we have the following set of simplectic variables defined by $\xi^{(1)}=(U_k, p_k; A_k, \pi_k, A_0, \pi_0; \Sigma_k, \phi_k ; \lambda_1 , \lambda_2)$, with the new canonical one-form defined by
\ba\label{19}
&&{}^{U}a^{(0)}_k =-p_k ; \,\, {}^{\Sigma}a^{(0)}_k =\phi_k;\,\, {}^{A}a^{(0)}_k =-\pi_k;\,\no
&&{}^{A_0}a^{(0)} =\pi_0;\,\,{}^{\lambda_1}a^{(0)} ={}^{1}\Omega;\,\,{}^{\lambda_2}a^{(0)} ={}^{2}\Omega.
\ea

\ni Hence, the first iterated symplectic matrix is written as 
\be\label{20}
f^{(1)}_{ij}(\vec x, \vec y) =
\begin{pmatrix}
{\bf A}_{ij} & {\bf B}_{j,y} \cr -{\bf B}_{i,x}^T & {\bf G}_{ij} \cr 
\end{pmatrix} \delta(\vec x -\vec y)
\ee

\ni where
\ba\label{21}
&&{\bf A}_{ij}=
\begin{pmatrix}
0 & \delta_{ij} & 0 & 0 & 0 \cr -\delta_{ij} & 0 & 0 & 0 & 0 \cr 0 & 0 & 0 & -\delta_{ij} & 0 \cr 0 & 0 & \delta_{ij} & 0 & 0 \cr 0 & 0 & 0 & 0 & 0 \cr
\end{pmatrix} , \no
&&{\bf B}_{j,y}=
\begin{pmatrix}
0 & 0 & 0 & 0 & 0 \cr 0 & 0 & 0 & \partial_j^y & 0 \cr 0 & 0 & 0 & 0 & 0 \cr 0 & 0 & 0 & 0 & 0 \cr -1 & 0 & 0 & -g\partial_y^2 & 0 \cr
\end{pmatrix}, \no
&&{\bf G}_{ij}=
\begin{pmatrix}
0 & 0 & 0 & 0 & 1 \cr 0 & 0 & -\delta_{ij} & g\partial_j^y & 0 \cr 0 & \delta_{ij} & 0 & 0 & -\partial_j^y \cr 0 & -g\partial_i^x & 0 & 0 & 0\cr -1 & 0 & \partial_i^x & 0 & 0 \cr
\end{pmatrix} \,\,,
\ea

\ni and we can see that $f^{(1)}_{ij}$ is a singular matrix. From this result, we can determine its zero-mode as being
\be\label{22}
\bar{\nu}_{\alpha} =(\bar{\nu}_{i}^{U}, {\bf 0}, {\bf 0}, {\bf 0}, \bar{\nu}^{A_0}, 0, \bar{\nu}_{i}^{\Sigma}, \bar{\nu}_{i}^{\phi}, \bar{\nu}^{\lambda_1}, \bar{\nu}^{\lambda_2} )\,\,,
\ee

\ni where $\bar{\nu}_{i}^{U}=\partial_i \bar{\nu}^{\lambda_1}; \;\; \bar{\nu}_{i}^{\Sigma}=\partial_i \bar{\nu}^{\lambda_2}; \;\; \bar{\nu}_{i}^{\phi}=-g\partial_i \bar{\nu}^{\lambda_1}; \;\; \bar{\nu}^{A_0}=-\bar{\nu}^{\lambda_2}$ and $\bar{\nu}^{\lambda_1}$, $\bar{\nu}^{\lambda_2}$ are arbitrary functions. Thus, from this zero-mode in Eq. (\ref{22}) we have that
\ba\label{24}
&&{}^{3}\Omega =\int d^3\vec x \,\Big[\bar\nu_{i}^{U}(\vec x){{\delta}\over{\delta U_i(\vec x)}}+\bar\nu_{i}^{\Sigma}(\vec x){{\delta}\over{\delta \Sigma_i(\vec x)}} \no
&+&\bar\nu^{A_0}(\vec x){{\delta}\over{\delta A_0(\vec x)}}+ \bar\nu_{i}^{\phi}(\vec x){{\delta}\over{\delta \phi_i(\vec x)}}\Big]\int d^3\vec y \,V^{(1)}(\vec y)  \nonumber\\
&=&\int d^3\vec x \,\bar\nu_{i}^{\lambda_2}(\vec x)[\nabla .\vec \pi (\vec x)]=0\,\,.
\ea

Once again, as $\bar\nu^{\lambda_2}$ is an arbitrary function, we obtain a new set of constraint relations given by
\be\label{25}
{}^{3}\Omega=\nabla \cdot \vec \pi =0\,\,.
\ee

Now, following the FJ method, the second-iterated Lagrangian can be written as
\be\label{26}
{\cal L}^{(2)} = -\vec{p}.\dot{\vec{U}} +\vec{\phi}.\dot{\vec \Sigma} +\pi_\mu \dot{A}^{\mu} + \dot{\lambda}_1 {}^{1}\Omega + \dot{\lambda}_2 {}^{2}\Omega +\dot{\lambda}_3 {}^{3}\Omega  - V^{(2)}\,\,,
\ee

\ni where 
\be\label{27}
V^{(2)} = V^{(1)}\Big|_{{}^{3}\Omega=0} =V^{(1)}\,\,.
\ee

\ni From the above Lagrangian we can find the following canonical non-zero one-form 
\ba\label{28}
&&{}^{U}a^{(0)}_k =-p_k ; \,\, {}^{\Sigma}a^{(0)}_k =\phi_k;\,\, {}^{A}a^{(0)}_k =-\pi_k;\,\, \\
&&{}^{A_0}a^{(0)} =\pi_0;\:{}^{\lambda_1}a^{(0)} ={}^{1}\Omega;\,\,{}^{\lambda_2}a^{(0)} ={}^{2}\Omega; \,\, {}^{\lambda_3}a^{(0)} ={}^{3}\Omega\,\,, \nonumber
\ea

\ni which leads to the corresponding third-iterated symplectic matrix,
\be\label{29}
f^{(2)}_{ij}(\vec x, \vec y) =
\begin{pmatrix}
{\bf A}_{ij} & {\bf \bar B}_{j,y} \cr -{\bf \bar B}_{i,x}^T  & {\bf \bar G}_{ij} \cr 
\end{pmatrix} \delta(\vec x -\vec y)\,\,
\ee

\ni where ${\bf A}_{ij}$ has the same expression given in (\ref{21}), and
\ba\label{30}
{\bf \bar B}_{j,y}=
\begin{pmatrix}
0 & 0 & 0 & 0 & 0 & 0 \cr 0 & 0 & 0 & \partial_j^y & 0 & 0 \cr 0 & 0 & 0 & 0 & 0 & 0 \cr 0 & 0 & 0 & 0 & 0 & \partial_j^y \cr -1 & 0 & 0 & -g\partial_y^2 & 0 &0 \cr
\end{pmatrix}, \no
{\bf \bar G}_{ij}=
\begin{pmatrix}
0 & 0 & 0 & 0 & 1 & 0 \cr 0 & 0 & -\delta_{ij} & g\partial_j^y & 0 & 0  \cr 0 & \delta_{ij} & 0 & 0 & -\partial_j^y & 0 \cr 0 & -g\partial_i^x & 0 & 0 & 0 & 0\cr -1 & 0 & \partial_i^x & 0 & 0 & 0 \cr 0 & 0 & 0& 0 & 0 & 0 \cr
\end{pmatrix}\,\,,
\ea

\ni and once again, we can see that $f^{(2)}$ is singular and the zero-mode associated with this matrix is
\be\label{32}
{\bar{\bar\nu}}_\alpha =(\bar{\nu}_{\alpha}, \bar{\bar{\nu}}^{\lambda_3})\,\,,
\ee
where $\bar{\nu}_{\alpha}$ has the same expression given by Eq. (\ref{22}). However, the zero-mode ${\bar{\bar\nu}}_\alpha$ generates the constraint  ${}^{3}\Omega$ again, the zero-mode does not generate any new constraints and, consequently, the symplectic matrix remains  singular. It characterizes the theory as a gauge theory.

In order to obtain a regular symplectic matrix a gauge fixing term must be added to the theory. The choice of this condition can be suggested by many reasons, the most important being the simplification that it may introduce in the theory. In the Maxwell theory, the condition usually employed to gauge fixing is the Coulomb gauge
\be\label{33}
A_0 =0\;\;, \quad \nabla \cdot \vec A = 0\,\,.
\ee

\ni However, concerning the theory described by the Lagrangian in (\ref{01}), where a charged compressible fluid is immersed in an electromagnetic field, the condition (\ref{33}) is not sufficient to promote a gauge fixing, to do that we need an extra condition. In this case, an appropriate choice is the  ``Lorentz Gauge" to a compressible fluid \cite{albert1}, where
\be \label{35}
\partial_\alpha U^{\alpha} =0
\quad \mbox{or} \quad
\nabla \cdot \vec U + \Gamma_0 =0 \,\,,
\ee
which is directly related to the condition relative to the compressibility of the fluid \cite{albert1}.  Thus, considering Eqs. (\ref{33}) and (\ref{35}) with gauge fixing conditions
\be\label{37}
{}^{4}\bar{\Omega} = \nabla \cdot \vec A, \quad \mbox{and} \quad
{}^{5}\bar{\Omega} = \nabla \cdot \vec U +\Gamma_0 \,\,,
\ee

\ni we them obtain a new Lagrangian density
\ba\label{38}
&&{\cal L}^{(3)} = -\vec{p}\cdot\dot{\vec{U}} +\vec{\phi}\cdot\dot{\vec \Sigma} -\vec\pi \cdot \dot{\vec A} + \dot{\lambda}_1 (\nabla \cdot \vec p \no
&+&g\nabla \cdot \vec\Sigma) + \dot{\lambda}_2 (\pi_0 -\nabla \cdot\vec\phi) +\dot{\lambda}_3 (\nabla \cdot\vec\pi) +\dot\lambda_4 (\nabla \cdot\vec A) \no
&+& \dot\lambda_5 (\nabla \cdot\vec U +\Gamma_0)  - V^{(3)} \,\,,
\ea

\ni where 
\ba\label{39}
&&V^{(3)} = -\vec{\pi}\cdot\vec{\Sigma} -{1\over2}{\vec p}^2  - {1\over2}(\nabla \times \vec{U})^2 \no
&&- {1\over2}(1+g^2)\vec \Sigma^2 +{1\over 2}(1+g^2)\vec A \cdot(\nabla^2 \vec A) \no
&&-\vec{\phi} \cdot (\nabla^2 \vec A)\,\,
\ea

\ni is associated with the symplectic variables $\xi^{(3)}=(U_k, p_k; A_k, \pi_k,  \pi_0; \Sigma_k, \phi_k ; \lambda_1 , \lambda_2, \lambda_3 ,\lambda_4 , \lambda_5)$. From the expression for the potential $V^{(3)}$ we can see what theory's dynamical variables are.  They are part of the canonical set $(p_k , U_k )$, $(\pi_k , A_k )$ and $(\phi_k ,\Sigma_k )$. The new canonical one-form is defined by
\ba\label{40}
&&{}^{U}a^{(0)}_k =-p_k ; \,\, {}^{\Sigma}a^{(0)}_k =\phi_k;\,\, {}^{A}a^{(0)}_k =-\pi_k;\,\,\\
&&{}^{\lambda_1}a^{(0)} =\partial_i  p_i +g\partial_i \Sigma_i;\,\,{}^{\lambda_2}a^{(0)} =\pi_0 -\partial_i \phi_i; \nonumber\\
&&{}^{\lambda_3}a^{(0)} =\partial_i \pi_i ;\;\;{}^{\lambda_4}a^{(0)} =\partial_i A_i;\; {}^{\lambda_5}a^{(0)}=\partial_i U_i +\Gamma_0. \nonumber
\ea

\ni These relations can lead us to the corresponding third-iterated symplectic matrix
\be\label{41}
f^{(3)}_{ij}(\vec x, \vec y) =
\begin{pmatrix}
{\bf \tilde A}_{ij} & {\bf \tilde B}_{ij,y} \cr {-\bf \tilde B}_{ji, x}^T & {\bf \tilde G}_{ij}\cr 
\end{pmatrix} \delta(\vec x -\vec y)
\ee

\ni where
\ba\label{42}
{\bf \tilde A}_{ij}&=&
\begin{pmatrix}
0 & \delta_{ij} & 0 & 0 & 0 & 0 \cr -\delta_{ij} & 0 & 0 & 0 & 0 & 0 \cr 0 & 0 & 0 & -\delta_{ij} & 0 & 0 \cr 0 & 0 & \delta_{ij} & 0 & 0 & 0 \cr 0 & 0 & 0 & 0 & 0 & 0 \cr 0 & 0 & 0 & 0 & 0 & 0 \cr
\end{pmatrix} , \no
{\bf \tilde B}_{ij,y}&=&
\begin{pmatrix}
0 & 0 & 0 & 0 & 0 & \partial_j^y \cr 0 &\partial_j^y & 0 & 0 & 0 & 0  \cr 0 & 0 & 0 & 0& \partial_j^y & 0 \cr 0 & 0 & 0 &\partial_j^y& 0 & 0 \cr 0 & 0 & 1 & 0 & 0 & 0 \cr -\delta_{ij} & g\partial_j^y & 0 & 0 & 0 & 0 \cr
\end{pmatrix}, \no
{\bf \tilde G}_{ij}&=&
\begin{pmatrix}
0 & 0 & -\partial_j^y& 0 & 0 & 0 \cr 0 & 0 & 0 & 0 & 0 & 0 \cr \partial_i^x & 0 & 0 & 0 & 0 & 0 \cr 0 & 0 & 0 & 0 & 0 & 0 \cr 0 & 0 & 0 & 0 & 0 & 0 \cr 0 & 0 & 0 & 0 & 0 & 0 \cr
\end{pmatrix}.
\ea

\ni We can observe that $f^{(3)}_{ij}$ is not singular, therefore we can construct its inverse. The inverse of $f^{(3)}_{ij}$ is called the symplectic tensor
\begin{widetext}
\be\label{43}
(f^{(3)}_{ij})^{-1} = \nonumber
\ee
\be\label{44}
\left(\begin{array}{cccccccccccc}
0 & \delta_{ij}- {{\partial_i \partial_j}\over{\nabla^2}} & 0 & 0 & 0 & 0 & 0 & 0 & 0 & 0 & 0 & {{\partial_i}\over{\nabla^2}} \\
-\delta_{ij} + {{\partial_i \partial_j}\over{\nabla^2}} & 0 & 0 & 0 & g\partial_i & 0 & g\delta_{ij} & {{\partial_i}\over{\nabla^2}} & 0 & 0 & 0 & 0 \\ 
0  & 0 & 0 & -\delta_{ij} + {{\partial_i \partial_j}\over{\nabla^2}} & 0 & 0 & 0 & 0 & 0 & 0 & {{\partial_i}\over{\nabla^2}} & 0 \\
0 & 0 & \delta_{ij}- {{\partial_i \partial_j}\over{\nabla^2}} & 0 & 0 & 0 & 0 & 0 & 0 & {{\partial_i}\over{\nabla^2}} & 0 & 0 \\
0 & -g\partial_j & 0 & 0 & 0 & \partial_i & 0 & 0 & -1 & 0 & 0 & -g \\
0 & 0 & 0 & 0 & -\partial_j & 0 & \delta_{ij} & 0 & 0 & 0 & 0 & {{\partial_i}\over{\nabla^2}}\\
0 & -g\delta_{ij} & 0 & 0 & 0 & -\delta_{ij} & 0 & 0 & 0 & 0 & 0 & g{{\partial_i}\over{\nabla^2}}\\
0 & -{{\partial_j}\over{\nabla^2}}& 0 & 0 & 0 & 0 & 0 & 0 & 0 & 0 & 0 & -{{1}\over{\nabla^2}}\\
0 & 0 & 0 & 0 & 1 & 0 & 0 & 0 & 0 & 0 & 0 & 0\\
0 & 0 & 0 & -{{\partial_j}\over{\nabla^2}} & 0 & 0 & 0 & 0 & 0 & 0 & {{1}\over{\nabla^2}} & 0\\
0 & 0 & -{{\partial_j}\over{\nabla^2}}& 0 & 0 & 0 & 0 & 0 & 0 & -{{1}\over{\nabla^2}} & 0 & 0\\
-{{\partial_j}\over{\nabla^2}}& 0 & 0 & 0 & g & 0 & -g{{\partial_j}\over{\nabla^2}} & {{1}\over{\nabla^2}} & 0 & 0 & 0 & 0 
\end{array}\right)\delta^{(3)}(\vec x - \vec y).
\ee
\end{widetext}
Moreover, we can relate $\lambda_1 =U_0$, $\lambda_2 =\Sigma_0$ and $\lambda_3 = A_0$.  In this way, from (\ref{44}) it is possible to identify the following FJ's generalized brackets given by
\bigskip
\ba\label{46}
&&\{ A_i(\vec x), \pi_j (\vec y) \} = \left(-\delta_{ij} + {{\partial_i \partial_j}\over{\nabla^2}}\right)\delta(\vec x -\vec y) \no
&&\{\Sigma_i (\vec x), \phi_j (\vec y)\} = \delta_{ij}\delta(\vec x -\vec y)\,\,, \no
&&\{ U_i(\vec x), p_j (\vec y) \} = \left(\delta_{ij}- {{\partial_i \partial_j}\over{\nabla^2}} \right)\delta(\vec x -\vec y)
\ea 
\ni and
\ba \label{47}
&&\{p_i(\vec x), \phi_j (\vec y)\} = g\delta_{ij}\delta(\vec x - \vec y)\,\,, \no
&&\{p_i(\vec x), \pi_0 (\vec y)\} = g\partial_{i}\delta(\vec x - \vec y)\,\,,
\ea

\ni where the Dirac brackets for the electromagnetic fields correspond to the one we have found in preview works \cite{albert1,albert0}, as well as the Dirac brackets for the higher-derivative terms in the electromagnetic fields, $\Sigma$ and $\phi$.  Besides, we have found the Dirac brackets for the dynamic variables of the compressible fluid $p$ and $U$, the last of Eqs.(\ref{46}). Finally, as a result of the coupling between the charged compressible fluid and the electromagnetic field we found two Dirac brackets between the fluid and electromagnetic fields, Eqs. (\ref{47}), which are both zero when there is no coupling between them.

\section{Conclusions.}

In this paper we have analyzed the gauge invariance of the theory which describes the charged compressible fluid interacting with an electromagnetic field (\ref{01}) by using the Faddeev-Jackiw method. We have found the constraints, the gauge transformations and we have obtained the generalized FJ brackets.
\acknowledgments

The authors thank CNPq (Conselho Nacional de Desenvolvimento Cient\' ifico e Tecnol\'ogico), Brazilian scientific support federal agency, for partial financial support, Grants numbers 302155/2015-5, 302156/2015-1 and 442369/2014-0 and E.M.C.A. thanks the hospitality of Theoretical Physics Department at Federal University of Rio de Janeiro (UFRJ), where part of this work was carried out.

\end{document}